\documentclass[prb,twocolumn,superscriptaddress,amsmath,amssymb,showpacs,floatfix,preprintnumbers]{revtex4-1}
\usepackage{amsmath}
\usepackage{amssymb}
\usepackage{graphicx}
\usepackage[export]{adjustbox}
\usepackage{epstopdf}
\usepackage{color}
\usepackage{epsfig}
\usepackage{amsmath}
\usepackage{amssymb}

\DeclareMathAlphabet{\bi}{OML}{cmm}{b}{it}

\begin{document}

\title{Comment on "Impurity spectra of graphene \\ under electric and magnetic fields"}
\author{ R. Van Pottelberge}\email{robbe.vanpottelberge@uantwerpen.be}
\affiliation{Departement Fysica, Universiteit Antwerpen, Groenenborgerlaan 171, B-2020 Antwerpen, Belgium}

\author{M. Zarenia}\email{mohammad.zarenia@uantwerpen.be}
\affiliation{Departement Fysica, Universiteit Antwerpen, Groenenborgerlaan 171, B-2020 Antwerpen, Belgium}

\author{F. M. Peeters}\email{francois.peeters@uantwerpen.be}
\affiliation{Departement Fysica, Universiteit Antwerpen, Groenenborgerlaan 171, B-2020 Antwerpen, Belgium}

\begin{abstract}
In a recent publication [Phys. Rev. B \textbf{89}, 155403 (2014)], the authors investigated the spectrum of a Coulomb impurity in graphene in the presence of magnetic and electric fields using the coupled series expansion approach. In the first part of their publication they investigated how Coulomb impurity states collapse in the presence of a perpendicular magnetic field. We argue that the obtained spectrum does not give information about the atomic collapse and that their interpretation of the spectrum regarding atomic collapse is not correct. We also argue that the obtained results are only valid up to the dimensionless charge $\mid \alpha\mid=0.5$ and in order to obtain correct results for $\alpha>0.5$ a proper regularisation of the Coulomb interaction is required. Here we present the correct numerical results for the spectrum for arbitrary values of $\alpha$.
\end{abstract}

\maketitle
In Ref. [\onlinecite{sun}] Sun and Zhu investigated the spectrum of a Coulomb impurity in the presence of a perpendicular magnetic field as function of the strength of the Coulomb potential $\alpha$. They use a point size Coulomb potential and investigate both an attractive and repulsive Coulomb potential. For $\mid\alpha\mid>\mid j\mid$ ($j$ is the total angular quantum number) they found that the Landau Levels (LLs) suddenly disappear and bound states cannot exist (Fig. 2 in Ref. [\onlinecite{sun}]). They argued that the disappearance of these states is the signature of atomic collapse in the presence of a magnetic field. A symmetric spectrum was found between an attractive and repulsive potential, namely the spectrum is invariant under the transformations $\alpha\rightarrow -\alpha$ and $E\rightarrow -E$. 

However we argue that their results are only valid for $\mid \alpha\mid<0.5$. In this region one is able to obtain solutions for all angular quantum numbers. When the strength of the impurity exceeds $\mid\alpha\mid=0.5$, no bound states for the angular quantum numbers $l=0,-1$ are found as can be seen in Fig. 2 of Ref. [\onlinecite{sun}] (for higher charges LLs with higher angular quantum number also start to disapear). The fact that these LLs disapear is actually \textit{not} a signature of atomic collapse. This can be traced back to the singular nature of the point size Coulomb potential, causing a breakdown of the model for $\mid\alpha\mid>0.5$. In order to obtain physical solutions and to investigate the atomic collapse phenomena one needs to regularise the Coulomb potential in order to remove this singularity. This can be done by using a physically more realistic potential, for example by taking into account the finite size of the impurity or the fact that in some experiments the impurity is placed at a certain distance from the graphene sheet [\onlinecite{crommie}]. The fact that the results for a point size impurity are only valid when $\mid \alpha\mid<0.5$ has been previously stated by D.S. Novikov [\onlinecite{novikov}]. He investigated the scattering of electrons in the presence of a point size Coulomb impurity and also noted that the solutions are only valid and consistent when $\mid\alpha\mid<0.5$. This point is also supported by Ref. [\onlinecite{gusynin}] in which the authors note that the Hamiltonian is not self-adjoint for $\mid\alpha\mid>0.5$ without regularisation.
 
In order to obtain the full spectrum for impurity charges beyond 0.5 we solve the Dirac equation, see Eq. (2) in Ref. [\onlinecite{sun}], with the \textit{regularised} Coulomb potential [\onlinecite{sobol},\onlinecite{andrei}]:
\begin{equation}
V(r)=\frac{\alpha}{\sqrt{r^2+d^2}}.
\end{equation}
This regularised potential corresponds to the physical situation where an impurity is placed at a distance $d$ from the graphene sheet. We calculated the spectrum numerically using the finite elements method.

\begin{figure*}
\includegraphics[scale=0.45]{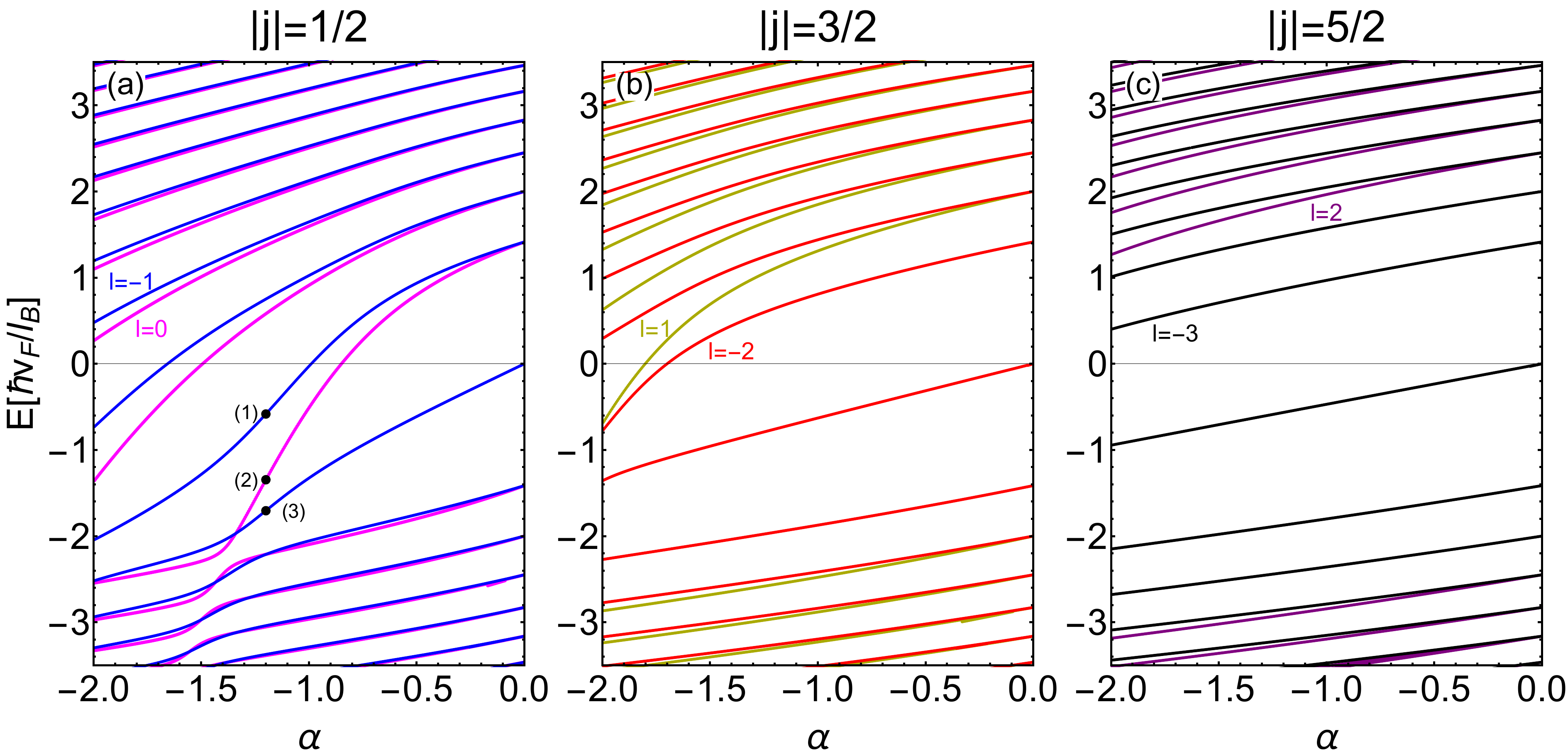}
\centering
\caption{Landau levels as function of the impurity strength for an attractive potential in the presence of a magnetic field $B=10$ T, which corresponds to a magnetic length of $l_B=8.1 \text{ nm}$. $l$ and $j=l+1/2$, respectively, indicate the angular momentum and total angular momentum quantum numbers. In each panel the levels with the same total angular momentum $\mid j\mid$ are displayed. We assumed the charge was placed at a distance $d=0.4$ nm from the graphene sheet.}
\end{figure*}

In Fig. 1 we plot the spectrum as function of the charge strength $\alpha$ for the same quantum numbers as done in Ref. [\onlinecite{sun}] and we used the same color coding for clarity as in Ref. [\onlinecite{sun}]. We only plot the spectrum of the attractive potential since there is symmetry between the spectrum of an attractive and repulsive potential. The LLs with the same total angular quantum number $\mid j\mid$ are plotted in the same panel as labelled. The $l$ quantum number is related to the $j$ quantum number as follows: $j=l+1/2$.

\begin{figure}
\includegraphics[scale=0.443]{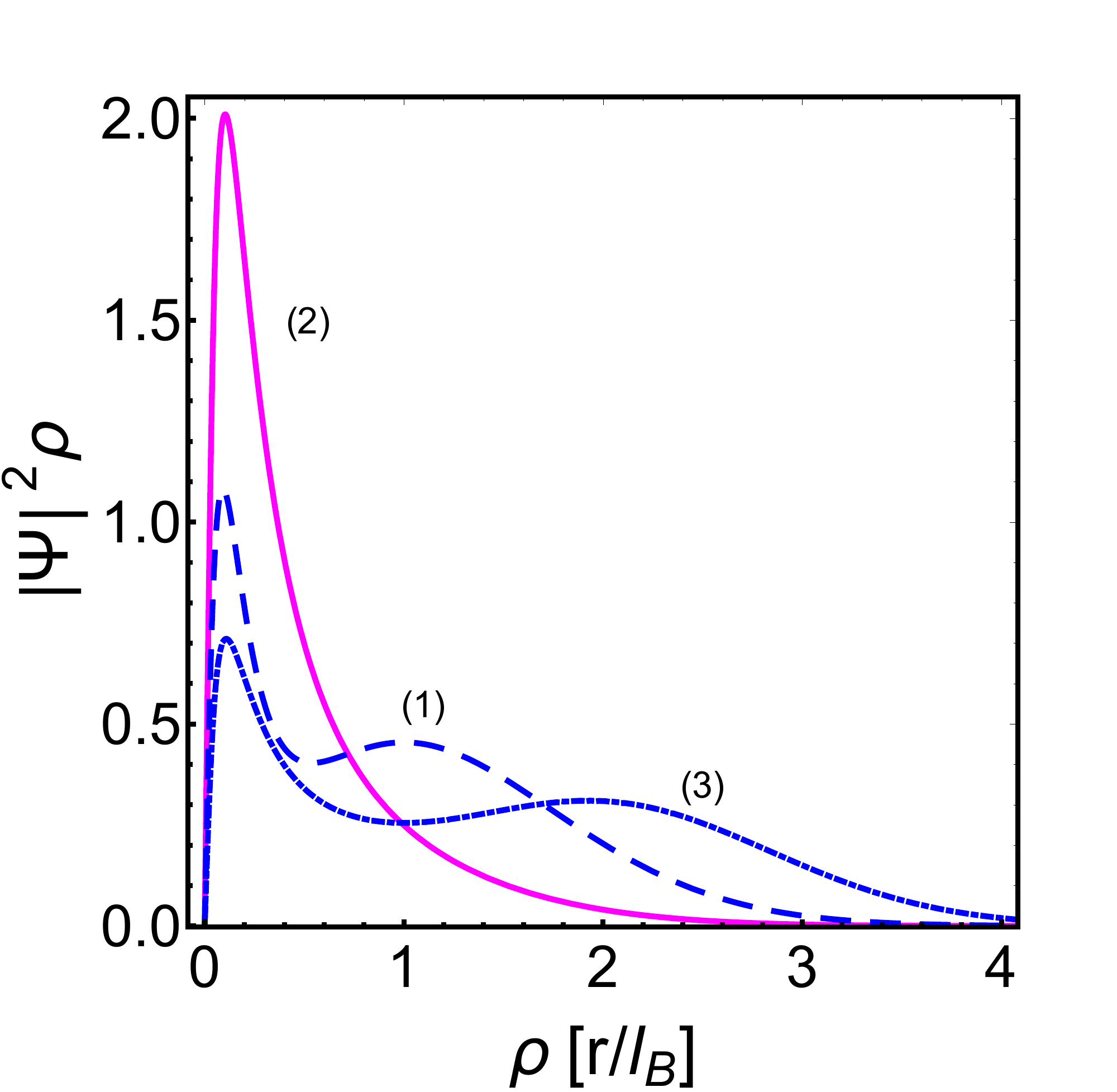}
\centering
\caption{Probability densities for the states indicated in Fig. 1(a) by a solid circle.}
\end{figure}

The states with $l=0,-1$ in Fig. 1(a) exhibit similar behaviour in the region $\mid \alpha\mid<0.5$, however, we find many more states for both quantum numbers than in Ref. [\onlinecite{sun}]. It seems that the states corresponding to higher LLs are not displayed in Fig. 2 of Ref. [\onlinecite{sun}]. As expected due to the regularisation we are able to find solutions for the levels $l=0,-1$ when $\mid\alpha\mid>0.5$. We see that once $\mid\alpha\mid$ exceeds 0.5 the lowest electron LLs for both $l=0$ and $l=-1$ dive sharply towards the negative hole region, and at the same time their probability density shows a sharp peak very close to the impurity. This can be clearly seen in Fig. 2 where the probability density is plotted for the solid circles indicated in Fig. 1(a). The lowest electron LL moves through the band of negative LLs exhibiting a series of anticrossings. The signature of atomic collapse is thus not seen in the disappearance of the levels. However, anticrossings are formed in the negative hole region, the lowest electron LL dives sharply into the negative hole region, and the probability density starts to show a very sharp peak near the impurity which are clear signatures of atomic collapse. One can define the critical charge in the presence of a magnetic field as the charge for which
the first electron LL anticrosses with the first hole LL, which gives $\mid\alpha\mid\approx 1.4$ [\onlinecite{sobol}]. In Figs. 1(b, c) we show the LLs for the total angular quantum number $j=1/2$ and $j=3/2$, respectively. We find similar results for charges up to $\mid \alpha\mid=\mid j\mid$. Due to the regularisation we were able to find solutions for $\mid\alpha\mid>\mid j\mid$. It can be clearly seen in Fig. 1(b) how the lowest electron states for both quantum numbers dive into the negative hole region. Thus collapse does not appear as the disappearance of these levels. Our conclusions are supported by a recent numerical study using the tight binding Hamiltonian [\onlinecite{dean}].
\\

In this comment we have studied an attractive Coulomb potential in the presence of a perpendicular magnetic field. We argued that the results obtained by Sun and Zhu in Ref. [\onlinecite{sun}] for the case of a point size Coulomb potential are only valid for $\mid \alpha\mid<0.5$, where solutions can be obtained for all the angular quantum numbers. In order to obtain all the LLs for $\mid\alpha\mid>0.5$ a regularisation of the Coulomb potential is required. We performed this regularisation and successfully obtained the LLs when $\mid\alpha\mid>\mid j\mid$. We showed that the atomic collapse does not manifest itself through the disappearance of energy states, as claimed by Sun and Zhu, but rather in: 1) a sudden decrease of the lowest electron LL in the hole region, 2) sharp peak of the probability distribution very close to the impurity, and 3) a series of anticrossings in the negative hole region. We also showed that not all LLs were shown, in the paper by Sun and Zhu.
\\

We thank Matthias Van der Donck for fruitful discussions. This work was supported by the Flemish Science Foundation (FWO-Vl) and the Methusalem funding of the Flemish Government.


\begin{thebibliography}{99}
\bibitem{sun} Songyang Sun and Jia-Lin Zhu, Phys. Rev. B \textbf{89}, 155403 (2014).
\bibitem{crommie} Yang Wang, Dillon Wong, Andrey V. Shytov, Victor W. Brar, Sangkook Choi, Qiong Wu1, Hsin-Zon Tsai, William Regan, Alex Zettl, Roland K. Kawakami, Steven G. Louie, Leonid S. Levitov, Michael F. Crommie, Science \textbf{340}, 6133 (2013).
\bibitem{novikov} D. S. Novikov, Phys. Rev. B \textbf{76}, 245435 (2007). 
\bibitem{gusynin} O. V. Gamayun, E. V. Gorbar, and V. P. Gusynin, Phys. Rev. B \textbf{83}, 235104 (2011). 
\bibitem{sobol} O. O. Sobol, P. K. Pyatkovskiy, E. V. Gorbar, and V. P. Gusynin Phys. Rev. B \textbf{94}, 115409(2016). 
\bibitem{andrei} Adina Luican-Mayer, Maxim Kharitonov, Guohong Li, Chih-Pin Lu, Ivan Skachko, Alem-Mar B. Goncalves, K. Watanabe, T. Taniguchi, and Eva Y. Andrei, Phys. Rev. Lett. \textbf{112}, 036804 (2014). 
\bibitem{dean} D. Moldovan, M. R. Masir, and F. M. Peeters, 2D Mater. \textbf{5}, 015017 (2018).
\end{thebibliography}
\end{document}